\documentclass{article}
\usepackage{spconf,amsmath,graphicx}
\usepackage{float}
\usepackage{graphicx}
\usepackage{url}
\usepackage{algorithm}
\usepackage{algpseudocode}
\usepackage{multirow}

\input{./Definitions}

\title{Adaptive Multi-Corpora Language Model Training for Speech Recognition}
\name{Yingyi Ma, Zhe Liu, Xuedong Zhang}
\address{Meta AI, Menlo Park, CA, USA}
\begin{document}
\ninept
\maketitle
\begin{abstract}
Neural network language model (NNLM) plays an essential role in automatic speech recognition (ASR) systems, especially in adaptation tasks when text-only data is available. In practice, an NNLM is typically trained on a combination of  data sampled from multiple corpora. Thus, the data sampling strategy is important to the adaptation performance. Most existing works focus on designing static sampling strategies. However, each corpus may show varying impacts at different NNLM training stages. In this paper, we introduce a novel adaptive multi-corpora training algorithm that dynamically learns and adjusts the sampling probability of each corpus along the training process. The algorithm is robust to corpora sizes and domain relevance. Compared with static sampling strategy baselines, the proposed approach yields remarkable improvement by achieving up to relative 7\% and 9\% word error rate (WER) reductions on in-domain and out-of-domain adaptation tasks, respectively.
\end{abstract}
\section{Introduction}
\label{sec:intro}
Language models have been commonly used to promote automatic speech recognition (ASR) performance through first-pass fusion \cite{KanWuNgu2018,KimShaMah2021,ChaJaiNav2016} or second-pass rescoring \cite{LiuWanChe2014,XuCheGao2018,LiPovKhu,IriZeyAlb2019}. Recent studies show that neural network based language model (NNLM) obtains significantly stronger performance than traditional $n$-gram models given its better capability of modeling long-range dependency \cite{MikKarMar2010,CheLiuGal2015,XuLiWan2018,GraFerSan2006,Gra2012}. 

Typically, an external language model can be trained separately from ASR models, thus bringing further flexibility to its choice of training data and training strategy. Since only text data is required for training any language model, it provides enormous advantages on leveraging external language models on adaptation tasks for ASR applications where there is a mismatch between  source and target domains.

The proper performance of a language model highly depends on the quality and quantity of training data in any adaptation task. However, it is hard to guarantee this especially when the target domain is lack of adaptation data \cite{Bel2004, li2020empirical}, or such data is not fully accessible due to privacy pursuit \cite{GanRasHof2018,LiuLiBak2021,GulBeaMot2021,CuiLuKin2021}.  Alternatively, we can resort to text data from multiple corpora. Some corpora may be in a similar domain with the target while some may not. Hence, how to improve the performance of a language model by effectively leveraging large-scale multiple corpora becomes a critical research topic.

Training a single $n$-gram language model on multiple corpora is relatively straightforward. This is usually done by training an $n$-gram model for each corpus and then interpolating them to form a final model. The interpolation weights are usually optimized over a validation set from the target domain. The linearity of the parameters in $n$-gram language models grants direct interpolation, while it is hard to see an easy extension to non-linear NNLMs. Instead of model interpolation, mixing data from diverse corpora fits better for mini-batch fashion training. However, we found that the strategies for sampling data from multiple corpora lack study. A few works explore data mixing methods based on the relevance between each corpus and the target domain \cite{RajFilTiw2019,SchGau2005}, nonetheless, all these methods employ static sampling strategies, where each corpus's sampling probability is precomputed and fixed along the training process. Recent studies show that the order of feeding corpus into training is important for boosting the model performance \cite{AgrSinSch2021,ChaYehDem2021,LiuLaiWon2020}, indicating that the sampling probability of each corpus may need to be adjusted dynamically. 

Motivated by the challenges and insights above, we propose a novel adaptive multi-corpora NNLM training algorithm. We adjust the sampling probabilities based on the estimated contribution that each corpus can devote in the following training stage. Specifically, at the beginning of each epoch, we fine-tune the current model on each corpus separately, then interpolate the fine-tuned models and optimize the interpolation weights over the validation dataset of the target domain. The optimized weights will then be leveraged as the sampling probabilities over multiple corpora in the current training epoch. Our method can work with any arbitrary sized corpora and adaptively take into account the importance of each corpus along the entire training process. We conduct experiments on in-domain and out-of-domain adaptation scenarios, with results indicating that the proposed adaptive sampling probability adjustment is effective in improving the performance of NNLM for ASR applications.
\section{Related Works}
Although various works have studied data sampling and weighting strategies across language model and deep learning training tasks \cite{FerDow2018, RenZenYan2018, ShuXieYi2019}, most of them are applied to the sample level other than the corpus or dataset level. Learning to reweight samples usually requires the computation of second-order derivatives in every backward pass, which leads to high complexity and thus could be heavy-loaded for multi-corpora training. 

As another line of research, recent works have shown that the order of corpora plays a critical role in training language models \cite{AgrSinSch2021,ChaYehDem2021,LiuLaiWon2020}. Hence, abundant metrics for measuring the relevance over corpora have been designed, and curriculum learning approaches in lieu of data mixing strategies have also been proposed. However, a well-designed adaptive data mixing strategy naturally incorporates the importance of corpora order during the training process.


Towards joint training with multiple corpora, multi-task based training strategies are introduced \cite{TusIriSch2016,dSaIllFoh2022,LiuHeChe2019}. These works train a multi-head language model with several shared hidden layers and corpora-specific output layers. In adaptation tasks, they learn the interpolation weights for combining corpora-specific layers and result in a single NNLM. However, the parameter size and inference latency of such model would grow with the number of corpora, making this type of approaches less attractive in practical applications. 

Inspired by the interpolation of $n$-gram models, authors in \cite{RajFilTiw2019} propose to train an $n$-gram model on each corpus and optimize their interpolation weights over a validation set. Then during the NNLM training, minibatches are stochastically constructed by drawing samples from each corpus with probability according to the interpolation weights. These weights are only learned once from $n$-gram models and fixed over the entire NNLM training process. Unlike this approach, our proposed method adaptively optimizes the sampling probability per corpus during the training process. 


\section{Methods}
Consider an NNLM adaptation task where we are given $K$ different training corpora $D_1,\ldots,D_K$ with each $D_k = \{x_i^{(k)}\}_{i=1}^{N_k}$ consisting of $N_k$ records, as well as a validation corpus $D_{\tau} = \{y_i\}_{i=1}^{N_{\tau}}$ from the target domain. The goal is to train a model with records sampled from the $K$ corpora that can achieve optimized performance on the target domain. Note that $N_{\tau}$ is usually a much smaller number compared with these ${N_k}$'s, and the target domain could be close-in-domain with a few ones of these $K$ corpora or a completely different domain.




Towards this goal, it is crucial to develop a sampling strategy for mixing multiple corpora during model training. We propose to adjust the sampling probability per corpus to fit best for each training stage and optimize it over the validation set dynamically. 
 Specifically, the NNLM training is divided into an upper-level task and a lower-level task. Optimizing the data sampling probability across multiple corpora  is regarded as the upper-level task, while the lower-level task is in charge of learning model parameters of NNLM given current sampling probabilities. The upper-level task and lower-level task are conducted in an alternating manner until model convergence.

We first introduce a generic training framework using mixed data from multiple corpora in the following subsection.

\subsection{Mixed Data Training}

Define $\Dcal = \Mcal(D_1,\ldots, D_K| W)$ as a mixed data distribution over corpora $D_1,\ldots, D_K$ such that
\begin{align}
\label{formula:mix}
\forall x \sim \Dcal:\;&P(x \in D_k) = w_k, \\
&P(x = x_i^{(k)}| x \in D_k)=\frac{1}{N_k},\;\forall k\in [1..K]
\end{align}
where $W = (w_1, \ldots, w_K)$ is the sampling probability over corpora and $\Sigma_{k=1}^Kw_k=1$. This mixed data distribution will sample data from corpus $D_k$ with probability $w_k$. That is, at each training iteration (i.e.~model update), we sample a minibatch of training records from $\Dcal$ with roughly $w_k$ portion of data coming from corpus $D_k$.


Given the mixed data distribution $\Dcal$, an NNLM $\theta$ can be trained by minimizing the negative log-likelihood of samples: 
\begin{align}
\label{formula:loss}
 \Lcal_{train}(\theta | \Dcal) = -\Sigma_{x \sim \Dcal} \log P_{\theta}(x)
\end{align}
where $P_{\theta}(x)$ is the model estimated probability of observing $x$ as a product of each token's probability given its preceding tokens.

As each corpus may show varying impacts on the target domain at different training stages, we will adjust the sampling probability $W$ over multiple corpora before each training epoch of NNLM, with details described in the next subsection.

\subsection{Sampling Probability Optimization}
Adjusting the sampling probability $W$ aims to optimize the contribution of each corpus in the following training. Thus, we first need to measure the effects of continuing training with each corpus, and then adapt the sampling probability accordingly. To this end, we propose to fine-tune the current NNLM using each corpus solely, after which interpolate $K$ fine-tuned models and optimize the interpolation weights over the validation set of the target domain. The learned weights will serve as the sampling probability in the next. We break down this process into the two steps below.

\subsubsection{The fine-tuning step}
Define a fine-tuning operator as 
\begin{align}
FT(\theta, D, S) \mapsto \theta_{FT}
\end{align}
which fine-tunes input model $\theta$ on corpus $D$ for $S$ iterations.
To fairly measure the contribution each corpus could devote in the following training process, we fine-tune the current model $\theta$ on each corpus for the same number of iterations $S$. Then, we will obtain $K$ different models $\theta_1,\ldots, \theta_K$:
\begin{align}
\theta_k = FT(\theta, D_k, S),\,\forall k \in [1..K]
\end{align}
The fine-tuning step is a continued training from the current model $\theta$ per corpus and conducted at the beginning of each epoch. 

\subsubsection{The interpolation weight optimization step}
Consider an interpolation weight optimization operator: 
\begin{align}
WO(\Theta, D_{\tau}) \mapsto W
\end{align}
where $\Theta = (\theta_1,\ldots, \theta_K)$ is a collection of $K$ fine-tuned models. The $WO$ operator will optimize the model interpolation weights $W = (w_1,\ldots, w_K)$ over the performance on $D_{\tau}$ and then output the optimized weights. 
Specifically, the $WO$ operator can be defined as follows:
\begin{align}
W= &\argmin_{w_1,\ldots, w_K} -\Sigma_{y_i \in D_{\tau}}\log(\Sigma_{k=1}^K w_k\cdot P_{\theta_k}(y_i))
\\ &\,s.t.\quad w_k \in [0,1],\,\forall k \in [1..K]
\\ &\qquad\;\;\Sigma_{k=1}^K w_k = 1
\end{align}
By substituting $w_K$ with $1-w_1-\ldots -w_{K-1}$, the optimization problem above can be solved through either gradient descent (adopted in this work) or an expectation-maximization (EM) algorithm.

As the interpolation weights well reflect the importance of each corpus in the following training process, we can make use of them as the sampling probabilities over corpora for the current NNLM training epoch (\ref{formula:mix})-(\ref{formula:loss}).

\subsection{The Adaptive Multi-Corpora Training Framework}
To chain all the steps we have introduced, in Algorithm \ref{alg:main}, we present the adaptive sampling framework for multi-corpora NNLM training. The proposed method adaptively determines the sampling probability per corpus during the training process. Higher probabilities are assigned to the ones who will likely play more important roles towards improving the performance on target domain in certain training epochs, and vice versa. In the scenarios where the order of feeding corpora into training is vital, the proposed algorithm can automatically choose their relative orders by assigning higher sampling probability to a corpus when its turn is coming.

\begin{algorithm}
\caption{Adaptive multi-corpora training}
\label{alg:main}
\begin{algorithmic}
\State Input: $K$ training corpora $D_1,\ldots, D_K$, validation dataset of target domain $D_{\tau}$, number of fine-tuning iterations $S$
\State Initialize NNLM $\theta^0$
\For{epoch $t = 1,2,\ldots, T$}
\For{each corpus $k=1,2,\ldots, K$}
\State $\theta_k^t = FT(\theta^{t-1}, D_k, S)$
\EndFor
\State $W^t = WO((\theta_1^t,\ldots, \theta_K^t), D_{\tau})$
\State Construct $\Dcal^t = \Mcal(D_1, \ldots, D_K|W^t)$
\State Learn $\theta^t$ through training on $\Dcal^t$ for one epoch:
\State $\argmin_{\theta^t}\left(-\Sigma_{x \sim \Dcal^t} \log P_{\theta^t}(x)\right)$
\EndFor
\State Output: $\theta^T$
\end{algorithmic}
\end{algorithm}

Oftentimes, corpora available for training could be of diverse sizes due to data scarcity. The proposed algorithm is unfastidious in the various sizes of corpora. It performs the same number of training iterations when fine-tuning on each corpus, and if some corpora are of petite size but turn out to be frequently sampled, it simply reflects the importance of these corpora in certain training stages.


In Algorithm \ref{alg:main}, the frequency of optimizing sampling probabilities is once per training epoch. It can be flexibly adjusted in different  applications in practice. For example, we can adapt the sampling probabilities every $Q$ training iterations instead and the rest components in the algorithm should naturally follow.

It is noteworthy that the proposed method works differently than the bi-level optimization framework \cite{JenFav2018,LiuGaoZha2021} in the sense that the sampling probabilities learned at the end of training process only represents its optimality for the final training stage and can not be regarded as optimal over the entire training process. The proposed algorithm dynamically picks the optimal sampling probabilities at each training stage. 


Since the proposed method requires fine-tuning on each corpus before each training epoch to determine the sampling probabilities over corpora, it needs additional $K \cdot S\cdot T$ training iterations than the conventional NNLM training with static sampling probabilities. Notice that the fine-tuning process on each corpus are independent with each other and thus can be conducted in parallel for better efficiency.
\section{Experiments}
\subsection{Datasets}
Table~\ref{table:corpora} summarizes the corpora used in our experiments as well as their train, validation, and test splits. Among them, we have
\begin{itemize}
\item Publicly available datasets, including Fisher speech corpus (\textsl{fisher}), Wall Street Journal corpus (\textsl{wsj}), and Wikitext-103 (\textsl{wiki}). For \textsl{fisher} corpus, we only utilize text training data in this study and each record represents a conversation. For \textsl{wsj} corpus, we use \textit{nov93dev} as the validation set and \textit{nov92} as the test set;
\item In-house datasets, which contain video corpus sampled from public social media videos (\textsl{video}), and three conversational speech corpora with different topics (\textsl{conv1}, \textsl{conv2}, \textsl{conv3}) collected using mobile devices through crowd-sourcing from a data supplier for ASR. All these datasets are de-identified before transcription; all transcribers and researchers do not have access to any user-identifiable information.
\end{itemize}
\begin{table}[ht]
\centering
\caption{Summary of data splits from multiple corpora.}
\begin{tabular}{l|ccc}
\hline
  & \multicolumn{3}{c}{Splits} \\ \cline{2-4}
Corpora & \multicolumn{1}{c|}{\begin{tabular}[c]{@{}c@{}}Train\\ (\#text records)\end{tabular}} & \multicolumn{1}{c|}{\begin{tabular}[c]{@{}c@{}} {Validation} \\ (\#utts)\end{tabular}} & \begin{tabular}[c]{@{}c@{}}Test\\ (\#utts)\end{tabular} \\ \hline
\textsl{conv1} & \multicolumn{1}{c|}{138K} & \multicolumn{1}{c|}{1K} & 4K \\
\textsl{conv2} & \multicolumn{1}{c|}{2000K} & \multicolumn{1}{c|}{-} & - \\
\textsl{video} & \multicolumn{1}{c|}{1100K} & \multicolumn{1}{c|}{-} & -   \\
\textsl{wiki}    & \multicolumn{1}{c|}{840K} & \multicolumn{1}{c|}{-} & -   \\
\textsl{fisher}  & \multicolumn{1}{c|}{12K} & \multicolumn{1}{c|}{-} & -   \\
\textsl{conv3} & \multicolumn{1}{c|}{-} & \multicolumn{1}{c|}{1K} & 12K  \\
\textsl{wsj} & \multicolumn{1}{c|}{-} & \multicolumn{1}{c|}{0.5K} & 0.3K  \\ \hline
\end{tabular}
\label{table:corpora}
\end{table}
\subsection{Setups}
We consider two adaptation scenarios in this study
\begin{itemize}
    \item \emph{In-domain adaptation}, where one of the training corpora is in the same domain with the target;
    \item \emph{Out-of-domain adaptation}, none of the training corpora being in the same domain with the target.
\end{itemize}

For each setting, NNLMs are trained on multiple corpora and integrated with an ASR model via first-pass shallow fusion. We then evaluate the performance of trained NNLMs on the test sets of target domain in terms of perplexity (PPL) and word error rate (WER).


Each NNLM contains an embedding layer of dimension 300, and 2 LSTM layers with 1500 hidden nodes, which has around 15 million parameters in total. The ASR is a RNN-T model with the Emformer encoder \cite{emformer2021streaming}, LSTM predictor, and a joiner. It has around 80 million parameters and is trained from scratch using the train split of LibriSpeech ASR corpus \cite{panayotov2015librispeech}.

The following multi-corpora NNLM training methods are compared in our experiments:
\begin{itemize}
  \item \texttt{uniform-weight}, which assigns the same sampling probability to each training corpus. Note that this method is close to the ``data merging'' approach where the models are trained on merged corpora, but simply merging data altogether fails to balance the size of each training corpus;
  \item \texttt{ngram-opt-weight}, the method presented in \cite{RajFilTiw2019}, where an $n$-gram model is trained on each corpus, and the optimized interpolation weights (with respect to the validation set) from these $n$-gram models are used as the fixed sampling probabilities over multiple corpora during NNLM training;
  \item \texttt{adaptive-weight}, our proposed method in Algorithm~\ref{alg:main}.
\end{itemize}


\subsection{Results}
\subsubsection{In-domain adaptation}
We first conduct a set of experiments on learning NNLMs with the train split of one corpus (\textsl{conv1}), two corpora (\textsl{conv1+conv2}), or three corpora (\textsl{conv1+conv2+video}). The validation and test sets of \textsl{conv1} are considered as the ones from the target domain. Hence, these experiments are regarded as in-domain adaptation tasks since the train split of \textsl{conv1} also appears in the training corpora.

Table~\ref{table:in-domain} demonstrates the PPL and WER evaluation results on the test set of \textsl{conv1}. We can observe that the proposed adaptive training algorithm achieves the best performance in both scenarios of two corpora training and three corpora training. Compared with \texttt{uniform-weight} and \texttt{ngram-opt-weight} methods, our approach results in relative 3\%-5\% and 5\%-7\% WER reductions, respectively. It is also expected that leveraging more corpora in the training set generally improves the NNLM quality.


\begin{table}[ht]
\vspace{-0.3cm}
\centering
\caption{PPL and WER results on \textsl{conv1} test set.}
\begin{tabular}{l|l|cc}
\hline Train Corpora & NNLM Training Method & PPL  & WER \\ \hline
 \textsl{n/a} & \texttt{without-NNLM} &  -  & 24.60 \\ \hline
\textsl{conv1} & \texttt{uniform-weight} & 54.2 & 20.87 \\ \hline
\multirow{3}{*}{\begin{tabular}[c]{@{}l@{}}\textsl{conv1+conv2}\end{tabular}}   
    & \texttt{uniform-weight}     & 43.5 & 19.75 \\
    & \texttt{ngram-opt-weight}    & 48.5 & 20.15 \\
    & \texttt{adaptive-weight} & \textbf{38.8} & \textbf{18.82} \\ \hline
\multirow{3}{*}{\begin{tabular}[c]{@{}l@{}}\textsl{conv1+conv2}\\ \textsl{+video}\end{tabular}} 
    & \texttt{uniform-weight}     & 36.8 & 19.20 \\
    & \texttt{ngram-opt-weight}     &  43.8  &  19.63  \\
    & \texttt{adaptive-weight} & \textbf{32.9} & \textbf{18.65}  \\ \hline
\end{tabular}
\label{table:in-domain}
\end{table}

\subsubsection{Out-of-domain adaptation}
Here, NNLMs are trained on three corpora (\textsl{conv1+conv2+video}) or five corpora (\textsl{conv1+conv2+video+wiki+fisher}), and evaluated on two target domains, \textsl{conv3} and \textsl{wsj}. Note that each target domain is different from any of the domains in the training corpora. Thus, this study is considered as out-of-domain adaptation.

The PPL and WER evaluation results on the test sets of \textsl{conv3} and \textsl{wsj} are presented in Table~\ref{table:out-domain:conv3} and Table~\ref{table:out-domain:wsj}, respectively. Similar to our previous findings, the introduced \texttt{adaptive-weight} method outperforms all other methods consistently. Compared with \texttt{uniform-weight} approach, our method obtains relative 5\%-9\% WER reductions on \textsl{conv3} test set and 6\%-9\% WER reductions on \textsl{wsj} test set. Compared with \texttt{ngram-opt-weight} approach, our method achieves relative 3\% WER reductions on \textsl{conv3} test set and 2\%-5\% WER reductions on \textsl{wsj} test set.


\begin{table}[ht]
\centering
\caption{PPL and WER results on \textsl{conv3} test set.}
\begin{tabular}{l|l|cc}
\hline Train Corpora & NNLM Training Method & PPL  & WER \\ \hline
 \textsl{n/a} & \texttt{without-NNLM} &  -  & 24.79 \\ \hline
\multirow{3}{*}{\begin{tabular}[c]{@{}l@{}}\textsl{conv1+conv2}\\ \textsl{+video}\end{tabular}}  
    & \texttt{uniform-weight}     & 65.6 & 18.95 \\
    & \texttt{ngram-opt-weight}    & 60.2 & 18.55 \\
    & \texttt{adaptive-weight} & \textbf{49.7} & \textbf{17.98} \\ \hline
\multirow{3}{*}{\begin{tabular}[c]{@{}l@{}}\textsl{conv1+conv2+}\\ \textsl{video+wiki}\\ \textsl{+fisher}\end{tabular}} 
    & \texttt{uniform-weight}     & 63.2 & 18.92 \\
    & \texttt{ngram-opt-weight}     &  48.2  &  17.91   \\
    & \texttt{adaptive-weight} & \textbf{41.5} & \textbf{17.34}  \\ \hline
\end{tabular}
\label{table:out-domain:conv3}
\end{table}
\begin{table}[ht]
\centering
\caption{PPL and WER results on \textsl{wsj} test set.}
\begin{tabular}{l|l|cc}
\hline Train Corpora & NNLM Training Method & PPL  & WER \\ \hline
 \textsl{n/a} & \texttt{without-NNLM} &  -  & 10.96 \\ \hline
\multirow{3}{*}{\begin{tabular}[c]{@{}l@{}}\textsl{conv1+conv2}\\ \textsl{+video}\end{tabular}}  
    & \texttt{uniform-weight}     & 91.2 & 9.82 \\
    & \texttt{ngram-opt-weight}    & 80.8 & 9.39 \\
    & \texttt{adaptive-weight} & \textbf{65.2} & \textbf{8.95} \\ \hline
\multirow{3}{*}{\begin{tabular}[c]{@{}l@{}}\textsl{conv1+conv2+}\\ \textsl{video+wiki}\\ \textsl{+fisher}\end{tabular}} 
    & \texttt{uniform-weight}     & 78.2 & 9.44 \\
    & \texttt{ngram-opt-weight}     &  66.9  &  9.10   \\
    & \texttt{adaptive-weight} & \textbf{62.4} & \textbf{8.86}  \\ \hline
\end{tabular}
\label{table:out-domain:wsj}
\end{table}



\subsection{Analysis}
To provide more insights on how the proposed adaptive training algorithm works, we conduct additional analysis on \textsl{conv1} in-domain adaptation task with two training corpora (\textsl{conv1+conv2}), corresponding to row 4 to 6 in Table~\ref{table:in-domain}.

\begin{figure}[ht!]
  \begin{minipage}[c]{0.51\linewidth}
   \centering
    \includegraphics[width=\linewidth]{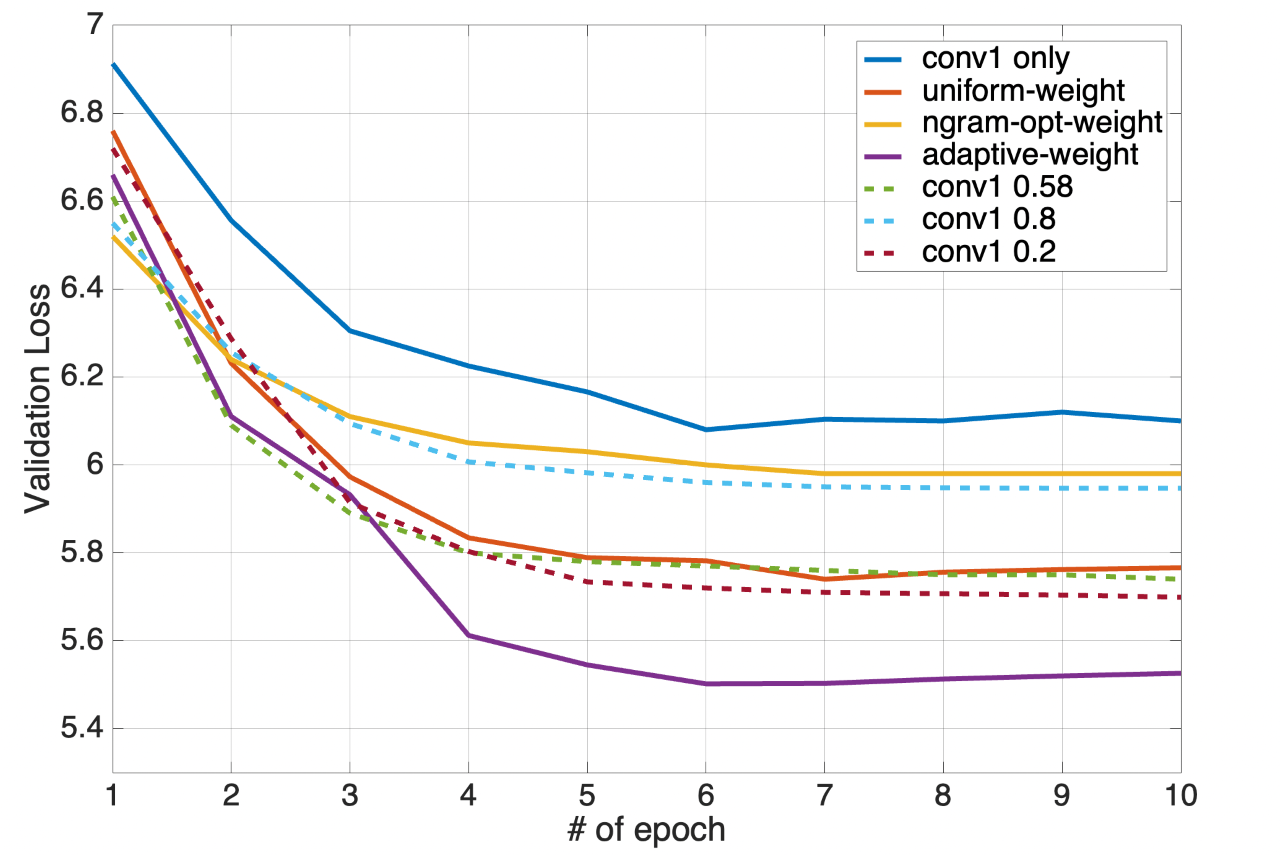} \\
    (a)
    \end{minipage}
  \hfill
  \begin{minipage}[c]{0.51\linewidth}
    \centering
    \includegraphics[width=\linewidth]{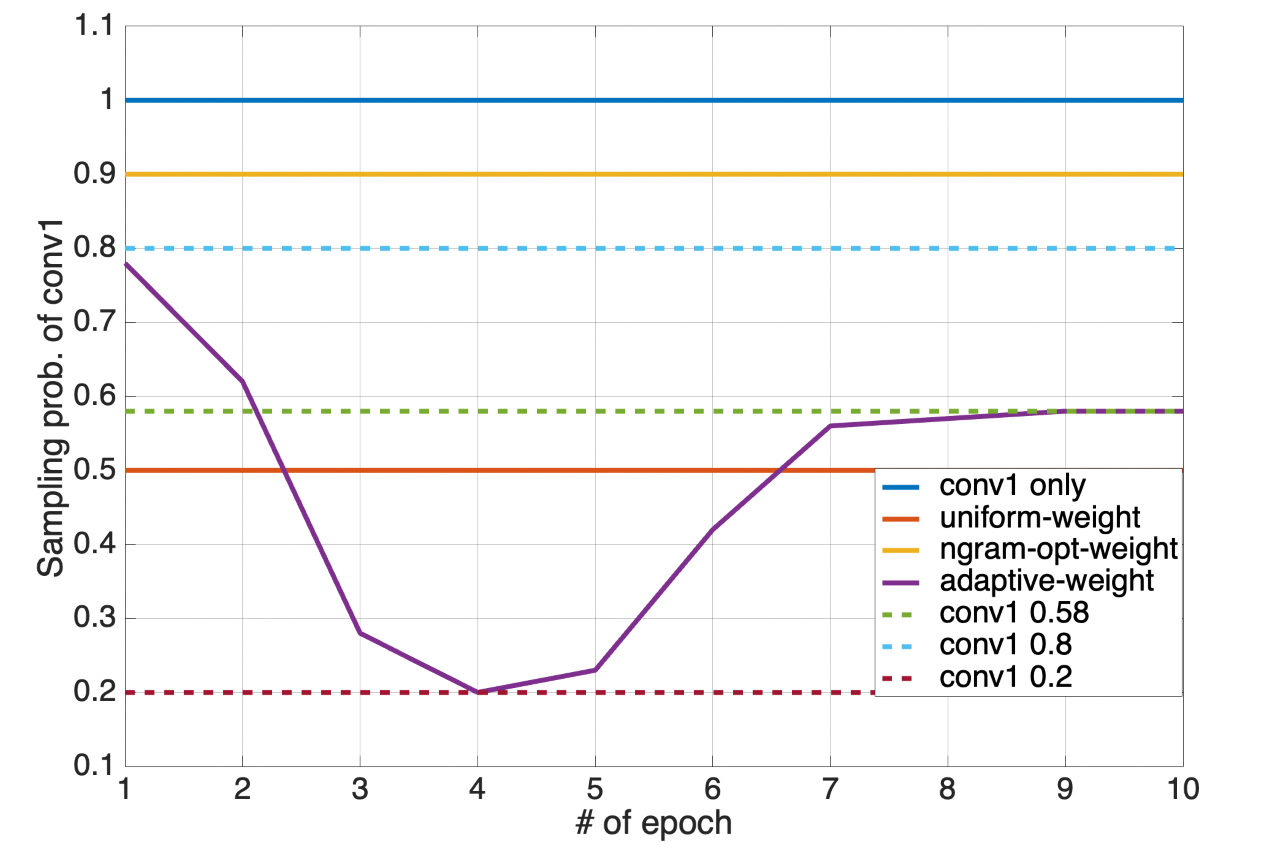} \\
    (b)
  \end{minipage}
  \caption{Training progress for various sampling strategies on \textsl{conv1} adaptation task; (a): validation loss versus training epoch; (b): sampling probability for \textsl{conv1} versus training epoch.}
  \label{fig:analysis}
\end{figure}

We present the training progress as follows: in Figure~\ref{fig:analysis}(a), we compare the validation loss along the training progress of multiple methods with different sampling strategies. Their corresponding sampling probability for the in-domain training corpus \textsl{conv1} is shown in Figure~\ref{fig:analysis}(b).

Seen from Figure~\ref{fig:analysis}(a), the proposed approach converges to a much lower validation loss. Without the use of out-of-domain data, training with in-domain data of \textsl{conv1} only performs the worst since the train split of \textsl{conv1} is relatively small. Similarly, leveraging the $n$-gram interpolation weights as the sampling probabilities makes it hard to perform well, because the $n$-gram model trained on in-domain adaptation set tends to receive a much higher interpolation weight. Assigning such a high probability to a small adaptation set of \textsl{conv1} results in insufficient use of the other corpus.

On the other hand, the proposed \texttt{adaptive-weight} method can implicitly take into account the size and importance of each corpus while adjusting the sampling probability during the training process. According to the sampling probability curve presented in Figure~\ref{fig:analysis}(b), its training process can be split into three stages. In the first two epochs, the model mainly focuses on learning from the in-domain adaptation set \textsl{conv1}. Starting from epoch three, continuing learning heavily from the in-domain corpus may lead to overfitting or early convergence. The model thus puts more efforts on learning from the out-of-domain corpus \textsl{conv2} until epoch six. After that, the learning rate becomes relatively smaller. The model then seeks a balanced weight for both corpora.

For more comparison, we train additional models by using the starting sampling probability (\textsl{conv1} 0.8), lowest  probability (\textsl{conv1} 0.2), or last probability (\textsl{conv1} 0.58) along the training progress of \texttt{adaptive-weight}, and assigning it as the static sampling probability during the training. However, none of these methods performs better or even close to the proposed method, further highlighting the significance of an adaptively adjustable data sampling strategy.

\section{Conclusion}
In this work, we introduce a novel adaptive multi-corpora training algorithm to improve the performance of NNLM for ASR applications. The proposed approach provides an adaptive data sampling strategy to effectively leverage each corpus along the training process. Experiment results show prominent WER improvement for both in-domain and out-of-domain adaptation tasks. Future work might include extending the presented multi-corpora algorithm to end-to-end ASR model training.

\vfill\pagebreak
\bibliographystyle{IEEEbib}
\footnotesize
\bibliography{refs}

\end{document}